\documentclass[12pt]{article}

\usepackage{amsmath,amsfonts,epsfig,mathrsfs,todonotes,yfonts}
\usepackage{xcolor}
\usepackage{cite}
\usepackage{stmaryrd}
\usepackage{mdframed}
\usepackage{scalerel}
\usepackage[top=2.5cm, bottom=2.5cm, left=2.75cm, right=2.75cm]{geometry} 
\usepackage{dsfont}
\numberwithin{equation}{section}
\usepackage{hyperref}
 \hypersetup{
     colorlinks=true,
     linkcolor=black,
     filecolor=blue,
     citecolor=blue,      
     urlcolor=cyan,
     }
\usepackage[most]{tcolorbox}

\newcommand{\re}[1] {(\ref{#1})}

\newcommand{\pa}{\partial} 

\newcommand{\al}{\alpha}
\newcommand{\be}{\beta}
\newcommand{\ga}{\gamma}

\newcommand{\de}{\delta}

\newcommand{\ber}{\begin{eqnarray}}
\newcommand{\eer}[1]{\label{#1}\end{eqnarray}}
\newcommand{\eero}{\end{eqnarray}}

\newcommand{\balg}{\begin{align}}
\newcommand{\ealg}{\end{align}}

\newcommand{\beq}{\begin{equation}}
\newcommand{\eeq}{\end{equation}}
\newcommand{\bea}{\begin{eqnarray}}
\newcommand{\eea}{\end{eqnarray}}

\newcommand{\rd}[1]{{\color{red}{#1}}}

\newcommand{\nn}{\nonumber}
\newcommand{\na}{\nabla}

\newcommand{\half}{{\textstyle{\frac12}}}

\newcommand{\quart}{{\textstyle{\frac14}}}

\def\HollowBox #1#2{{\dimen0=#1 \advance\dimen0 by -#2
       \dimen1=#1 \advance\dimen1 by #2
        \vrule height #1 depth #2 width #2
        \vrule height 0pt depth #2 width #1
        \llap{\vrule height #1 depth -\dimen0 width \dimen1} 
       \hskip -#2
       \vrule height #1 depth #2 width #2}}
\def\BOX{\HollowBox{.100in}{.010in}}

\usepackage{fancyhdr} 
\newcommand{\auth}{\large Ulf Lindstr\"om ${}^{a,b}$\footnote{email: ulf.lindstrom@physics.uu.se}
and {\"O}zg{\"u}r Sar{\i}o\u{g}lu ${}^a$\footnote{email: sarioglu@metu.edu.tr}}
\begin{document}
\begin{flushright}
{\small UUITP-48/21}\\
\vskip 1.5 cm
\end{flushright}

\begin{center}
{\Large{\bf Killing-Yano Cotton Currents}}
\vspace{.75cm}

\auth
\end{center}
\vspace{.5cm}
\vspace{.5cm}
\centerline{${}^a${\it \small Department of Physics, Faculty of Arts and Sciences,}}
\centerline{{\it \small Middle East Technical University, 06800, Ankara, Turkey}}
\vspace{.5cm}
\centerline{${}^b${\it \small Department of Physics and Astronomy, Theoretical Physics, Uppsala University}}
\centerline{{\it \small SE-751 20 Uppsala, Sweden}}

\vspace{1cm}


\centerline{{\bf Abstract}}
We discuss conserved currents constructed from the Cotton tensor and (conformal) 
Killing-Yano tensors (KYTs). We consider the corresponding charges generally and then 
exemplify with the four-dimensional Pleba\'nski-Demia\'nski metric where they are 
proportional to the sum of the squares of the electric and the magnetic charges. As part 
of the derivation, we {also} find the two {conformal} Killing-Yano tensors 
{of the Pleba\'nski-Demia\'nski metric} in the recently introduced coordinates 
of Podolsky and Vratny. The construction of asymptotic charges for the Cotton current 
is elucidated and compared to the three-dimensional construction in Topologically 
Massive Gravity. For the three-dimensional case, we also give a conformal superspace 
multiplet that contains the Cotton current in the bosonic sector. In a mathematical section,
we derive potentials for the currents, find identities for conformal KYTs and for KYTs in 
torsionful backgrounds.

\bigskip

\noindent

\vskip .5cm
  
\vspace{0.5cm}
\small
\pagebreak
\tableofcontents

\renewcommand{\thefootnote}{\arabic{footnote}}
\setcounter{footnote}{0}

\section{Introduction}
Killing-Yano tensors (KYTs) have long been studied. They play a role as square roots of 
second rank Killing tensors (KTs). In  gravity, supergravity and string theory they are used 
to find hidden symmetries for particles and backgrounds, to separate variables in 
Hamilton-Jacobi equations and to study the symmetries of the Dirac equation and its 
super extensions. 

General background on KTs and KYTs and their applications can be found in 
\cite{Frolov:2017kze,Chervonyi:2015ima} and \cite{Santillan:2011sh}. Some interesting 
aspects are: Finding new supersymmetries, covered in ``Susy in the sky'' \cite{Gibbons:1993ap} 
and their role in string theory covered, e.g. in \cite{DeJonghe:1996fb}. There are applications 
in General Relativity (GR) \cite{Carter,Walker:1970un} to $G$-structures 
\cite{Papadopoulos:2007gf, Papadopoulos:2011cz}, to WZW models \cite{Lunin:2020mxj}, 
to classical mechanics \cite{Cariglia:2014dfa} and to symmetries of the Dirac operator and 
super Laplacians \cite{Cariglia:2012ci,Howe:2016iqw}. Supersymmetric conformal KTs and 
KYTs are discussed in \cite{Howe:2015bdd}, in \cite{Howe:2018lwu} and \cite{Kuzenko:2020www}. 
Finally KTs arise in the context of hyperK\"ahler geometry \cite{Lindstrom:2009afn}.

The present paper grew out of the results in \cite{Lindstrom:2021qrk} which, in turn, was 
inspired by a result of Kastor and Traschen \cite{Kastor:2004jk,Kastor:2007tg}. These authors 
found a conserved current for an arbitrary rank KYT and it was used in certain backgrounds 
to construct asymptotic charges \cite{Olmez:2005by}. Here we continue and extend the discussion 
of conserved currents from \cite{Lindstrom:2021qrk}. In particular, we focus on a current constructed 
from the Cotton tensor and (conformal) KYTs. For this current we display the conserved 
charges in general, and for the four-dimensional ($4D$) case of the Pleba\'nski-Demia\'nski metric 
in particular. In the process we derive the two conformal KYTs (CKYTs) needed in the new 
Podolsky-Vratny coordinates. We further discuss the special case of three dimensions ($3D$) 
at some length, finding obstructions to asymptotic charges, comparing to results in Topologically 
Massive Gravity (TMG) and generalizing our current to $3D$ conformal supergravity in 
superspace. Continuing our mathematical study of KYTs, we finally include a section with 
identities for CKYTs.

The paper is organized as follows: In Sec.~\!\ref{CC} we define the Cotton current for an 
arbitrary dimension $D \geq 3$ and in Sec.~\!\ref{PlebDem} we study it in $4D$ using the 
Pleba\'nski-Demia\'nski metric as an example. In Sec.~\!\ref{3D} we descend to $3D$ and 
initiate a program for studying super KY currents in supergravity by generalizing the Cotton 
current to a supermultiplet in $3D$ conformal supergravity. Here we also discuss linearization 
around a background carrying a second rank KYT and compare some results to the ones derived 
in \cite{Deser:2003vh}. Sec.~\!\ref{math} presents the potentials for the Cotton currents.
The derivation of these involve some new mathematical relations for KYTs 
and CKYTs. To these we add a useful relation for the case of backgrounds with totally 
skew torsion. Sec.~\!\ref{conc} contains our conclusions. A short appendix 
has the conventions for our spinor algebra.

The setting of our discussions is GR on a manifold ${\cal M}$ of dimension $D$ that 
carries a metric $g$ and a rank $n$ KYT $f$, or CKYT $k$. The coordinate indices are denoted by 
latin letters $a,b,\dots$. When we consider extension to $3D$ supergravity in Sec~\!\ref{super}, 
we use spinor indices and denote vector indices by pairs of these.

\section{The Cotton Current}
\label{CC}
In this section we introduce conserved currents constructed from the Cotton tensor and 
(C)KYTs. The Cotton tensor is defined in $D \geq 3$ dimensions as  
\ber
C_{abc} \equiv 2 (D-2) \nabla _{[c} S_{b]a} =
2 \nabla _{[c} R_{b]a} - \frac{1}{(D-1)} g_{a[b} \nabla _{c]} R \,,
\eer{Cotdef}
where $R_{ab}$ is the Ricci tensor, $R$ is the curvature scalar and $S_{ab}$ is the Schouten tensor. 
In $D>3$ dimensions, it is related to the Weyl tensor $W_{abcd}$ by
\ber
C_{bdc} = \frac{D-2}{D-3} \, \na_a W^a_{~bcd}~.
\eer{}
It is traceless on all index pairs and satisfies 
\ber\nn
&& C_{abc} = C_{a[bc]} \,, \\[1mm] \nn
&& C_{[abc]} = 0 \,, \\[1mm]
&& \na^a C_{abc} = 0~.
\eer{Ccond}

We shall be interested in combining the Cotton tensor with a (C)KYT to construct conserved 
currents. A KYT of rank $n$ is an $n$-form $f$ that satisfies
\ber
\na_a f_{b_1 \dots b_{n}} = \na_{[a} f_{b_1 \dots b_{n}]} \,,
\eer{}
or equivalently
\ber
\na_{(a} f_{b_1) \dots b_{n}}=0~.
\eer{}
Similarly, a CKYT of rank $n$ is an $n$-form $k$ that satisfies
\ber
\na_a k_{b_1 \dots b_{n}} = \na_{[a} k_{b_1 \dots b_{n}]} 
+ \frac{n}{D-n+1} \, g_{a[b_1} \, \bar{k}_{b_2 \dots b_{n}]} \,,
\eer{cKYT}
where
\ber
\bar k_{b_1 \dots b_{n-1}} \equiv \na_c k^c_{~b_1 \dots b_{n-1}}~.
\eer{}

Consider the following combinations of the Cotton tensor $C$, a rank-2 KYT $f$ and a 
rank-2 CKYT $k$: 
\ber
\tcboxmath{
j^a \equiv C^{abc} f_{bc} \qquad \mbox{and} \qquad J^a \equiv C^{abc} k_{bc} \,.}
\eer{CotCur}
Note that
\ber
\na_a j^a = (\na_a C^{abc}) f_{bc} + C^{abc} \na_a f_{bc} 
= 0 + C^{abc} \na_{[a} f_{bc]} = 0~,
\eer{CovC}
where the last two relations in \re{Ccond} have been invoked. Hence $j^a$ is covariantly conserved. 
Similarly, we find that $J^a$ is also covariantly conserved: There is an additional term in 
the calculation corresponding to \re{CovC}
\ber
\na_a J^a = 0 + \frac{2}{D-1} \, C^{abc} \, g_{a[b} \, \bar{k}_{c]} = 0~,
\eer{CovC2}
where the last equality follows from the fact that the Cotton tensor is traceless. 

Since both the Cotton tensor and the CKYT are related to conformal properties of the manifold, 
one may ask about the transformation properties of $J^a$ under Weyl rescalings of the metric 
\( g^{\prime}_{ab} = e^{C} g_{ab} \). The Cotton tensor transforms as
\ber
C^{\prime}_{abc} = C_{abc} - \half (D-2) ( \pa_{d} C) W^{d}{}_{abc} \,.
\eer{Ctf}
A second rank CKYT transforms as \cite{Chervonyi:2015ima}
\ber
k^{\prime}_{ab} = e^{3C/2} k_{ab} 
\quad \mbox{and} \quad 
\bar{k}^{\prime}_{c} = e^{C/2} \left( \bar{k}_{c} + \half (\pa_{a} C) k^{a}{}_{c} \right) \,.
\eer{ktf}
From \re{Ctf}  and \re{ktf} it follows that 1-form current $J_a = C_{abc} k^{bc}$ transforms as
\ber
J^{\prime}_{a} \equiv C^{\prime}_{abc} k^{\prime \, bc}
= e^{-C/2} \left( C_{abc} - \half (D-2) ( \pa_{d} C) W^{d}{}_{abc} \right) k^{bc} \,,
\eer{}
and the vector current $J^a$ as
\ber
J^{\prime \, a} = e^{-3C/2} \left( C^{a}{}_{bc} - \half (D-2) ( \pa_{d} C) W^{da}{}_{bc} \right) k^{bc} \,.
\eer{}
This is clearly not a conformal scaling in general. In $3D$ the Weyl tensor does not exist 
which means that the covariant current {\em scales with $e^{-C/2}$} and that the contravariant 
current {\em scales with $e^{-3C/2}$}. This holds in arbitrary $D$ for conformally flat metrics 
or for constant scalings $C$.

We emphasize that the Cotton current \re{CotCur} defined here is unrelated to the 
other currents discussed in \cite{Lindstrom:2021qrk}. In particular, the currents in 
\cite{Lindstrom:2021qrk,Kastor:2004jk} are at most second order in the derivatives of the metric 
while the Cotton current \re{CotCur} is of third order. The current in \cite{Kastor:2004jk} is defined 
for arbitrary rank $n$ KYT and reduces to the Einstein current for $n=1$, while our Cotton 
current \re{CotCur} is only defined for rank $2$. Moreover, we examine the Cotton current 
for $n=2$ in $3D$, its relation to the discussion in \cite{Deser:2003vh}  and its generalization 
to supergravity in Sec.~\!\ref{3D}. The current in \cite{Kastor:2004jk} on the other hand vanishes 
for $n=2$ in $3D$, since it is proportional to the Weyl tensor as shown in 
\cite{Lindstrom:2021qrk}.

\section{Cotton charges for the Pleba\'nski-Demia\'nski metric}
\label{PlebDem}
In this section we want to give a nontrivial example where the newly introduced Cotton 
current $J^a$ \re{CotCur} can be used for defining a conserved charge. For that purpose we 
resort to the well-known, non-vacuum solution of GR, the celebrated Pleba\'nski-Demia\'nski 
metric \cite{Plebanski:1976gy,Kubiznak:2007kh}
\ber
ds^2 =  \Omega^{2} \left( -\frac{Q(r) (d \tau -p^{2} d \sigma)^{2}}{r^{2}+p^{2}}
+ \frac{P(p) (d \tau + r^{2} d \sigma)^{2}}{r^{2}+p^{2}}
+ \frac{r^{2}+p^{2}}{P(p)} d p^{2} + \frac{r^{2}+p^{2}}{Q(r)} d r^{2} \right) \,,
\eer{PDmet}
which solves the cosmological Einstein-Maxwell equations \( G_{ab} = 2 T_{ab} - \Lambda g_{ab} \) 
in $D=4$ when the metric functions $Q(r)$ and $P(p)$ are chosen as
\ber\nn
&& Q(r) = k + e^{2} +g^{2} -2 m r + \epsilon r^{2} - 2 n r^{3} - \left(k + \Lambda / 3 \right) r^{4} \,, \\
&& P(p) = k + 2 n p - \epsilon p^{2} + 2 m p^{3} - \left( k + e^{2} + g^{2} + \Lambda / 3 \right) p^{4} \,,
\eer{QPcho}
the conformal factor is 
\ber
\Omega(p,r) = (1 - p r)^{-1} \,,
\eer{ConFac}
and the vector potential is
\ber
A_{a} dx^{a} = - \frac{1}{r^{2}+p^{2}} \Big( e r \left( d\tau - p^{2} d\sigma \right)
+ g p \left( d\tau + r^{2} d\sigma \right) \Big) \,.
\eer{VecPot}
Out of the seven ($m, n, e, g, \epsilon, k$ and $\Lambda$) arbitrary real parameters, $e$ and $g$
represent the electric and magnetic charges, respectively, whereas $\Lambda$ is the cosmological
constant \cite{Griffiths:2005qp}. As shown in \cite{Kubiznak:2007kh}, the Pleba\'nski-Demia\'nski 
metric \re{PDmet} (without using the special choices \re{QPcho}, \re{ConFac} and \re{VecPot}) 
has a rank-2 CKYT \re{CKYT}, which reads
\ber
k = \Omega^{3} \Big( p dr \wedge \left( d\tau - p^{2} d\sigma \right) 
+ r dp \wedge \left(d\tau + r^{2} d\sigma \right) \Big) \,.
\eer{PDckyt}
It turns out that its dual \( h \equiv *k \) is also a rank-2 CKYT \re{CKYT} given by 
\cite{Kubiznak:2007kh}
\ber
h = \Omega^{3} \Big( r dr \wedge \left( p^{2} d\sigma - d\tau \right) 
+ p dp \wedge \left(r^{2} d\sigma + d\tau \right) \Big) \,.
\eer{*PDckyt}

These ingredients can be used for calculating the Cotton tensor $C$ and the conformal Cotton 
currents $J_{k}$ (using \re{PDckyt}) and $J_{h}$ (using \re{*PDckyt}). For the special choices 
\re{QPcho}, \re{ConFac} and \re{VecPot}, we find that Cotton current vectors are
\ber
J_{k} = \frac{4 (e^{2}+g^{2}) (1-pr)^4}{\left(p^{2}+r^{2}\right)^3} 
\Big( 2 p^{2} r^{2} \partial_{\tau} + (r^{2} - p^{2}) \partial_{\sigma} \Big) \,,
\eer{PDJcur}
and
\ber
J_{h} = \frac{4 (e^{2}+g^{2}) (1-pr)^4}{\left(p^{2}+r^{2}\right)^3} 
\Big( (p^{2} - r^{2}) \partial_{\tau} + 2 \partial_{\sigma} \Big) \,.
\eer{*PDJcur}

It seems natural to interpret the coordinate $\tau$ as the global time function so that
\( t^{a} = (\partial_{\tau})^{a} \) is a vector field on the spacetime satisfying 
\( t^{a} \na_{a} \tau = 1 \). We may then foliate the spacetime \re{PDmet} by Cauchy surfaces
$\Sigma_{\tau}$ parametrized by $\tau$ such that the metric $g_{ab}$
\re{PDmet} induces a spatial metric $\gamma_{ab}$ on each $\Sigma_{\tau}$ as
\[ \gamma_{ab} = g_{ab} + n_a n_b \,, \]
where $n^{a}$ is the unit normal vector field to the hypersurfaces $\Sigma_{\tau}$:
\[ n^{a} = - \frac{1}{\Omega} \sqrt{\frac{p^2+r^2}{Q(r)-P(p)}} \, (\partial_{\tau})^{a} \,, \;\;
n_{a} dx^a = \frac{\Omega}{\sqrt{p^2+r^2}} \Big( \sqrt{Q(r)-P(p)} \, d \tau 
- \frac{r^2 P(p) + p^2 Q(r)}{\sqrt{Q(r)-P(p)}} d \sigma \Big) \,, \]
with \( n^{a} n_{a} = -1 \). The relevant piece of the volume element on 
the hypersurface $\Sigma_{\tau}$ is
\[ \sqrt{\gamma} = \frac{\left(p^2+r^2\right)^{3/2} \Omega^3}{\sqrt{Q(r)-P(p)}} \,,
\quad \mbox{with} \quad \gamma \equiv \det{\gamma_{ab}} \,. \]
These can be used in defining a conserved charge ${\cal Q}$ in the usual way as
\ber
{\cal Q} \equiv \int_{\Sigma_{\tau}} d^3 x \, \sqrt{\gamma} \, J^{a} \, n_{a} \,,
\eer{Qdef}
provided the metric functions satisfy \( P(p) > 0 \) and \( Q(r) > 0 \), as suggested by the form 
of \re{PDmet}, and $\tau$ can be thought of as the global time function as discussed. 
Using \re{PDJcur} and \re{*PDJcur}, we find explicitly that
\bea
{\cal Q}_{k} & = & 4 (e^2 + g^2) \int  dr dp d\sigma \, 
\frac{r^2 P(p)-p^2 Q(r)}{\left(p^2+r^2\right) (P(p)-Q(r))} \,, \label{PDQk} \\
{\cal Q}_{h} & = & 4 (e^2 + g^2) \int dr dp d\sigma \, 
\frac{P(p)+Q(r)}{\left(p^2+r^2\right) (P(p)-Q(r))} \,. \label{PDQh}
\eea

At this stage, it is worth emphasizing that our only aim is to give a ``proof of principle", i.e. 
that the Cotton currents $J_{k}$ and $J_{h}$ can be used to come up with conserved 
charges ${\cal Q}_{k}$ and  ${\cal Q}_{h}$ \'a la \re{Qdef}. So rather than going into the 
discussion of different cases\footnote{See e.g. the invaluable work \cite{Griffiths:2005qp} 
for that.} that arise from studying the roots of the quartic functions $P(p)$ and $Q(r)$ 
to determine the ranges of the coordinates $p$ and $r$ (that guarantee that \( P(p) > 0 \) and 
\( Q(r) > 0 \)), we assume that  the coordinate patch(es) have properly chosen ranges 
for the variables $p, r$ and $\sigma$ such that the relevant 3-dimensional 
integrals in ${\cal Q}_{k}$ and ${\cal Q}_{h}$ \re{Qdef} are convergent. Note that our 
Cotton charges are proportional to the sums of the squares of the physical electric and 
magnetic charges.

\subsection{An ``improved" version}
In a recent work \cite{Podolsky:2021zwr}, the Pleba\'nski-Demia\'nski metric 
(for the version with a vanishing cosmological constant) has been cast in a new
``improved" form where the metric functions $P$ and $Q$ have been factorized, and the
parameters have been redefined to make their physical meaning more accessible
(See the beautiful work \cite{Podolsky:2021zwr} for details). Here we want to examine 
how this new version plays along with the Cotton current we propose.

The new kid on the block reads \cite{Podolsky:2021zwr}
\bea
ds^{2} & = & \frac{1}{\Omega^{2}} \left( - \frac{Q}{\rho^{2}} \left[ dt - \left( a \sin^{2}{\theta} 
+ 4 \ell \sin^{2}{(\theta/2)} \right) d\varphi \right]^{2} 
+ \frac{\rho^{2}}{Q} dr^{2} \right. \nonumber \\
& & \left. \qquad + \frac{\rho^{2}}{P} d\theta^{2} + \frac{P}{\rho^{2}} \sin^{2}{\theta} 
\left[a \, dt - \left( r^{2} + (a + \ell)^{2} \right) d \varphi \right]^{2} \right) \,, \label{newPD}
\eea
where
\begin{eqnarray*}
\Omega & = & 1 - \frac{\alpha a}{a^{2} + \ell^{2}} r \, ( \ell + a \cos{\theta}) \,, \\
\rho^{2} & = & r^{2} + (\ell + a \cos{\theta})^{2} \,, \\
P(\theta) & = & \left( 1 - \frac{\alpha a}{a^{2} + \ell^{2}} r_{+} (\ell + a \cos{\theta}) \right) 
\left(1 - \frac{\alpha a}{a^{2} + \ell^{2}} r_{-} ( \ell + a \cos{\theta}) \right) \,, \\
Q(r) & = & \left( r - r_{+} \right) \left( r - r_{-} \right) 
\left( 1+ \alpha a \, \frac{a - \ell}{a^{2} + \ell^{2}} \, r \right) 
\left( 1 - \alpha a \, \frac{a + \ell}{a^{2} + \ell^{2}} \, r \right) \,,
\end{eqnarray*}
and the special roots of $Q(r)$, that determine the two black-hole horizons are located at
\ber
r_{\pm} \equiv m \pm \sqrt{ m^{2} + \ell^{2} - a^{2} - e^{2} - g^{2}} \,. 
\eer{roots}
There are now six (non-negative) real parameters: $e$ and $g$ are still the electric 
and magnetic charges, respectively, $m$ is the mass, $a$ is the ``Kerr-like rotation", $\ell$
is the NUT parameter and $\alpha$ is acceleration \cite{Podolsky:2021zwr}.
This metric solves the Einstein-Maxwell equations \( G_{ab} = 2 T_{ab} \) with the vector 
potential
\[ A_{a} dx^{a} = -\sqrt{e^2 + g^2} \, \frac{r}{\rho^2} \, \left[ dt - \left( a \sin^{2}{\theta} 
+ 4 \ell \sin^{2}{(\theta/2)} \right) d\varphi \right] \,. \]
The two rank-2 CKYTs of \re{newPD} are\footnote{To our knowledge, these are new but
presumably they can be obtained from the ones in \cite{Kubiznak:2007kh} by employing
judicious coordinate transformations.}
\bea
\tilde{k} & = & \frac{1}{\Omega^3} \Big( - r \sin{\theta} d\theta \wedge 
\left[a \, dt - \left( r^{2} + (a + \ell)^{2} \right) d \varphi \right] \Big. \nonumber \\
& & \qquad \Big. + ( \ell + a \cos{\theta}) dr \wedge \left[ dt - \left( a \sin^{2}{\theta} 
+ 4 \ell \sin^{2}{(\theta/2)} \right) d\varphi \right]  \Big) \,, \label{newk} \\
\tilde{h} & = & \frac{1}{\Omega^3} \Big( r \, dr \wedge \left[ dt - \left( a \sin^{2}{\theta} 
+ 4 \ell \sin^{2}{(\theta/2)} \right) d\varphi \right] \Big. \nonumber \\
& & \qquad \Big. + ( \ell + a \cos{\theta}) \sin{\theta} d\theta \wedge 
\left[a \, dt - \left( r^{2} + (a + \ell)^{2} \right) d \varphi \right] \Big)
\,. \label{newh}
\eea

Unfortunately though, the analogous calculations for determining the charge ${\cal Q}$ \re{Qdef} 
for the most general form of \re{newPD} with its CKYTs \re{newk} and \re{newh} turn 
out to be too complicated. Even though we have explicitly determined the analogous integrands 
as in \re{PDQk} and \re{PDQh}, they are best left undisplayed. To have a more manageable 
example, we set the ``Kerr-like rotation parameter" $a=0$ in the metric \re{newPD},
which also makes $\Omega=1$ and $P=1$ \cite{Podolsky:2021zwr}:
\ber
ds^{2} = - \frac{Q}{\rho^{2}} \left( dt - 4 \ell \sin^{2}(\theta/2) d\varphi \right)^{2} 
+ \frac{\rho^{2}}{Q} dr^{2} + \left( r^{2} + \ell^{2} \right) \left(d\theta^{2} + \sin^{2}{\theta} 
d\varphi^{2} \right) \,,
\eer{newPDa0}
where we now have
\begin{eqnarray*}
Q(r) & = & \left(r-r_{+}\right) \left(r-r_{-}\right) \,, \quad \mbox{with} \quad 
r_{\pm} \equiv m \pm \sqrt{ m^{2} + \ell^{2}  - e^{2} - g^{2}} \,, \\
\rho^{2} & = & r^{2}+ \ell^{2} \,.
\end{eqnarray*}
The vector potential becomes
\[  A_{a} dx^{a} = -\sqrt{e^2 + g^2} \, \frac{r}{\rho^2} \, 
\left( dt - 4 \ell \sin^{2}{(\theta/2)} d\varphi \right) \,, \]
and \re{newPDa0} describes the non-singular charged Taub-NUT spacetime, which 
is also asymptotically flat as $r \to \pm \infty$ away from the 
\( \theta = \pi \) axis \cite{Podolsky:2021zwr}.

It turns out that setting $a=0$ in $\tilde{k}$ \re{newk},
\ber
\hat{k} = r (r^{2} + \ell^{2}) \sin{\theta} d\theta \wedge d \varphi + \ell dr \wedge dt 
- 4 \ell^2 \sin^{2}{(\theta/2)} dr \wedge d\varphi  \,, 
\eer{newka0}
makes $\hat{k}$ \re{newka0} a KYT, not a CKYT, of \re{newPDa0}. So we can not 
use it to define a Cotton current $J^a$, as was done previously, but it can be used to define
the Cotton current $j^a$ \re{CotCur}. However, the $a=0$ version of $\tilde{h}$ \re{newh} is 
still a CKYT of \re{newPDa0}:
\ber
\hat{h} = r dr \wedge dt - 4 \ell r \sin^{2}{(\theta/2)} dr \wedge d\varphi 
- \ell (r^{2} + \ell^{2}) \sin{\theta} d\theta \wedge d \varphi \,.
\eer{newha0}
With these ingredients, we have the Cotton current vectors
\bea
j_{\hat{k}} & = & 0 \,, \quad \mbox{so that we trivially have} \;\; {\cal Q}_{\hat{k}} = 0 \,, \nonumber \\
J_{\hat{h}} & = & \frac{8 (e^{2}+g^{2}) \left(r-r_{+}\right) \left(r-r_{-}\right)}{\left(r^{2}+\ell^{2}\right)^3} 
\Big( - \partial_{t} + 4 \ell \sin^{2}{(\theta/2)} \partial_{\varphi} \Big) \,. \label{*PDJcura0}
\eea
Finally, we find the charge ${\cal Q}_{\hat{h}}$ to be\footnote{Here the ranges of the metric functions 
are \( r \in (-\infty, \infty) \), \(\theta \in [0,\pi]\) and \( \varphi \in [0, 2 \pi] \) \cite{Podolsky:2021zwr}.}
\ber
{\cal Q}_{\hat{h}} = 32 \pi (e^{2}+g^{2}) \int_{-\infty}^{\infty} \, 
\frac{dr}{r^{2}+\ell^{2}} = \frac{32 \pi^2}{\ell} (e^{2}+g^{2}) \,.
\eer{theQ}

\section{$3D$ results}
\label{3D}
The Cotton tensor is particularly relevant in $3D$ where it plays a role similar to that of the 
Weyl tensor in higher dimensions. In this section we discuss various properties of the Cotton 
current pertaining to $3D$.

\subsection{The Super Cotton Current in  $3D$ Conformal Supergravity}\label{super}
In this subsection we introduce the conformal supergravity extension of the Cotton current.

\subsubsection{$3D$ Conformal Superspace Supergravity}

In $3D$, ${\cal{N}}=1$ Conformal Superspace Conformal Supergravity is defined by 
the following algebra\footnote{See appendix \ref{Spinor} for conventions.}\cite{Butter:2013goa}
\ber\nn
&&\{\na_\al,\na_\be\}=2i\na_{\al\be} \,, \\[1mm]\nn
&&[\na_a,\na_\al]=\quart (\ga_a)_{\al}^{~\be}W_{\be\ga\de}K^{\ga\de} \,, \\[1mm]
&&[\na_a,\na_b]=-{\textstyle \frac i8}\epsilon_{abc}(\ga^c)^{\al\be}\na_\al W_{\be\ga\de}K^{\ga\de}
-\quart \epsilon_{abc}(\ga^c)^{\al\be}W_{\al\be\ga}S^\ga \,,
\eer{alg}
where $K$ is the generator of special conformal transformation and $S$ is the generator 
of $S$ supersymmetry. The whole super conformal group has been gauged and dilation 
curvature has been set to zero. The Lorentz generator $M$ and the translation generator 
$P$ give rise to the covariant derivatives. We shall need the properties of the super 
Cotton tensor $W$ \cite{Kuzenko:2012ew} 
\ber\nn
&&W_{\al\be\ga}=W_{(\al\be\ga)} \,, \\[1mm]\nn
&&\na^\al W_{\al\be\ga}=0 \,, \\[1mm]
&&K_aW_{\al\be\ga}=0 \,,
\eer{W}
where the last relation says that $W$ is a primary superfield.

From 
\ber
\na_\al\na_\be=i\na_{\al\be}+\half\epsilon_{\al\be}\na^\ga\na_\ga
\eer{}
and the fact that, on a primary field $[\na_{\al\be},\na_\ga ]=0$, it follows that
\ber
&&\na^2\na_\al=-\na_\al\na^2=2i\na^\ga\na_{\ga\al} \,, \\[1mm]\nn
&&\{\na^2,\na_\al\}=0~,~~~[\na^2,\na_\al]=4i\na^\ga\na_{\ga\al} \,, \\[1mm]\nn
&&\na^\ga\na_\al\na_\ga=0~.
\eer{}
We take a superconformal Killing supervector field $\xi$ to be given by
\ber
\xi=\xi^a\na_a+\xi^\al\na_\al~,
\eer{}
where $\xi^a$ is a primary and where the components are related by
\ber
\na_\al\xi_{\mu\nu}=4i\epsilon_{\al(\mu}\xi_{\nu)} \,.
\eer{}
It follows that
\ber\nn
&&\na^\al\xi^{\mu\nu}=-4i\epsilon^{\al(\mu}\xi^{\nu)} \,, \\[1mm]\nn
&&\na_{(\al}\xi_{\mu\nu)}=0 \,, \\[1mm]\nn
&&\na_{(\al}\xi_{\be)}=\quart \na^\ga_{~(\al}\xi_{\be)\ga} \,, \\[1mm]\nn
&&\na_{\ga}\xi^{\ga}={\textstyle \frac 1 3}\na_a\xi^a \,, \\[1mm]\nn
&&\xi^\al={\textstyle \frac i 6}(\ga_a)^{\al\be}\na_\be\xi^a \,, \\[1mm]\nn
&&\na^2\xi_\al=i{\textstyle \frac 2 3}\na^\ga_{~\al}\xi_\ga \,, \\[1mm]\nn
&&\na_{(a}\xi_{b)}={\textstyle \frac 1 3}\eta_{ab}\na_c\xi^c \,, \\[1mm]
&&\na^{\be\ga}\xi^\al=-{\textstyle \frac 2 3}\epsilon^{\al(\be}\na^{\ga)}_{~\sigma}\xi^\sigma \,,
\eer{}
where the next to last relation defines a conformal Killing vector and the last one a 
conformal Killing spinor.

\subsubsection{The Super Cotton Current}
From the last line in \re{alg} we see that 
\ber
C_{ab}=\na_\al W_{\be\ga\de}~~~ \makebox{and}~~~ W_{\be\ga\de}
\eer{}
will correspond to the Cotton and the Cottino tensors at the component level \cite{Butter:2013goa}. 
Indeed $C_{ab}$ has the properties of the Cotton-York tensor
\ber
\na_aC^a_{~b}=0~,~~~C_{ab}=C_{(ab)}~,~~~C^a_{~a}=0~.
\eer{}

We may now construct a supergravity version of our Cotton current $J$ in \re{CotCur}. 
To this end we define
\ber\nn
&&k_\al=W_{\al\be\ga}\xi^{\be\ga} \,, \\[1mm]
&&k_{\al\be}=\na_\al k_\be =(\na_{(\al} W_{\be)\ga\de})\xi^{\ga\de}+4iW_{\al\be\ga}\xi^{\ga} \,.
\eer{}
These satisfy
\ber\nn
&&\na^\al k_\al=0 \,, \\[1mm]
&&\na^{\al\be} k_{\al\be}=0~,
\eer{}
and the lowest component of the first part of $k_{\al\be}$ is the bosonic Cotton current $J$ 
in \re{CotCur}.

In a more covariant form we have
\ber
\tcboxmath{
(k^A)=(k^{\al\be},k^\al)}
\eer{}
with
\ber
\na_A k^A=0~.
\eer{}

\subsection{Currents from linearization?}
\label{Linear}
Here we want to examine whether we can extend the discussion in subsection 6.1 of 
\cite{Lindstrom:2021qrk} to the currents introduced in Sec.~\!\ \ref{CC}. So we linearize the 
metric around a background metric and assume that there exists a KYT asymptotically. If 
the linearized current defined with this KYT is still conserved, we may try to construct its 
asymptotic charges. For simplicity, consider only rank-2 KYTs in what follows. For brevity, 
let us start by reproducing some formulas that will be relevant for the ensuing discussion:
\ber
(\Gamma_{~ab}^{c})_L & = & \half \bar g^{ce} \big( \bar\na_a h_{be} + \bar\na_b h_{ae} 
- \bar\na_e h_{ab}\big) \,, \nn \\[1mm]
( R^{a}{}_{b} )_{L} & = & \frac{1}{2} \left( \bar{\na}^{c} \bar{\na}^{a} h_{bc} + 
\bar{\na}_{c} \bar{\na}_{b} h^{ac} - \bar{\na}^{a} \bar{\na}_{b} h - \bar{\BOX} h^{a}{}_{b} \right)
- h^{ac} \bar{R}_{bc} \,, \nn \\[1mm]
R_{L} & = & {\bar{\nabla}}_{a} {\bar{\nabla}}_{b} h^{ab} - \bar{\BOX} h - h^{ab} \bar{R}_{ab} \,.
\eer{Lin}
The Cotton current $j^{a}$ \re{CotCur}, and its linearization, is most relevant in $D=3$. This
is due to the special role played by the Cotton tensor in the field equations of TMG 
\cite{Deser:1982vy,Deser:1981wh}, so let us set $D=3$ in what follows and consider the 
linearized current
\ber
j^{a}_L \equiv 2 (\nabla _{c} S^{a}{}_{b})_L \bar{f}^{bc} \,.
\eer{jlin} 
Since
\[ (\nabla _{c} S^{a}{}_{b})_L  = \bar{\nabla} _{c} (S^{a}{}_{b})_L 
+ (\Gamma_{~cd}^{a})_L \bar{S}^{d}{}_{b}
- (\Gamma_{~cb}^{d})_L \bar{S}^{a}{}_{d} \,, \]
and 
\[ \bar{S}_{ab} \, \bar{f}^{ac} + \bar{S}^{ac} \, \bar{f}_{ab} = 0 \]
that follows from \( R_{ab} \, f^{ac} + R^{ac} \, f_{ab} = 0 \) \cite{Lindstrom:2021qrk}, we find
\ber
j^{a}_L = \bar{\nabla} _{c} \bar{\ell}^{ac} \,,
\eer{}
where \( \bar{\ell}^{ac} = 2 \bar{f}^{bc} (S^{a}{}_{b})_L \). We emphasize that this result
has been obtained by working with a generic background. Unfortunately though, we do not
have \( \bar{\ell}^{ac} = \bar{\ell}^{[ac]} \), or that \( \bar{\nabla} _{a} j^{a}_L = 0 \) even
when the background is maximally symmetric. So we cannot use $j^{a}$ to 
define asymptotic AD charges.

It is worth comparing this result with the relevant piece of the linearized current employed in 
\cite{Deser:2003vh} to define conserved gravitational charges in TMG. There the part
relevant to our discussion is defined through the linearization of the York tensor\footnote{We 
use the term ``York tensor" to denote the dual of the Cotton tensor, which is also called 
the ``Cotton tensor" or ``Cotton-York tensor" in the literature.} \( C^{ab} \), which is
symmetric, traceless and identically conserved, and defined as
\ber
C^{ab} = \epsilon^{acd} \nabla_{c} S_{d}{}^{b} \,,
\eer{YT}
and is the dual of the Cotton tensor, i.e. \( C^{b}{}_{cd} = \epsilon_{acd} C^{ab} \).
In \cite{Deser:2003vh}, it is shown that it is always possible to  write 
\ber
C^{ab}_{L} \bar{\xi}_{b} = \bar{\nabla} _{b} \big( \bar{\cal{F}}^{[ab]}(\bar{\xi}) \big) \,,
\eer{Yxi}
where \( \bar{\xi}^{b} \) is a background Killing vector, 
\( \bar{\Xi}^{a} \equiv \bar{\epsilon}^{abc} \bar{\nabla}_{b} \bar{\xi}_{c} \)
is another Killing vector constructed out of $\bar{\xi}$ and
\begin{eqnarray}
{\cal F}^{ab}(\bar{\xi}) & = & \half \Big( {\cal F}^{ab}_\mathrm{E}(\bar{\Xi})
+  \bar{\epsilon}^{cbd} \bar{\xi}_{c} \big( G_{d}\,^{a} \big)_{L}
+ \bar{\epsilon}^{abd} \bar{\xi}_{c} \big( G_{d}\,^{c} \big)_{L}
+ \bar{\epsilon}^{acd} \bar{\xi}_{c} \big( G_{d}\,^{b} \big)_{L} \Big) \,, \label{GCScharden} \\
{\cal F}^{ab}_\mathrm{E}(\bar{\Xi}) & = & \bar{\Xi}_{c} \bar{\nabla}^{[a} h^{b]c} 
+ \bar{\Xi}^{[b} \bar{\nabla}_{c} h^{a]c} 
+ h^{c[b} \bar{\nabla}_{c} \bar{\Xi}^{a]} + \bar{\Xi}^{[a} \bar{\nabla}^{b]} h 
+ \half h \bar{\nabla}^{[a} \bar{\Xi}^{b]} \,, \label{einchar} \\
\big( G^{a}\,_{b}\big)_{L} & = & \big( R^{a}\,_{b}\big)_{L} 
- \half \delta^{a}\,_{b} \, R_{L} - 2 \Lambda h^{a}\,_{b} \,, \nn \\
R_{ab}^{L} & = & \half \left( \bar{\nabla}^{c} \bar{\nabla}_{b} h_{ac} 
 + \bar{\nabla}^{c} \bar{\nabla}_{a} h_{bc} - \bar{\square} h_{ab}
 - \bar{\nabla}_{a} \bar{\nabla}_{b} h \right) \,, \nn \\
R_{L} & = & \bar{\nabla}^{a} \bar{\nabla}^{b} h_{ab} 
- \bar{\square} h - 2 \Lambda h  \,, \nn
\end{eqnarray}
where all contractions, raising and lowering of indices are performed with respect to the
maximally symmetric background $\bar{g}_{ab}$ for which 
\( \bar{R}_{ab} = 2 \Lambda \bar{g}_{ab} \) and \( \bar{R} = 6 \Lambda \), with $\Lambda < 0$, 
\( h_{ab} \equiv g_{ab} - \bar{g}_{ab} \) denotes the deviations from the background, 
\( h \equiv \bar{g}^{ab} h_{ab} \), $\bar{\nabla}$ indicates the covariant derivative with 
respect to the background and \( \bar{\square} \equiv \bar{\nabla}^{a} \bar{\nabla}_{a} \).

Since we have 
\( (\nabla_{c} S_{d}{}^{b})_{L} = \bar{\nabla} _{c} (S_{d}{}^{b})_{L} \) on such a background,
it follows that \( C^{ab}_{L} = \bar{\epsilon}^{acd} \bar{\nabla} _{c} (S_{d}{}^{b})_{L} \).
This leads to the conclusion that it is possible to write
\[ \bar{\epsilon}_{acd} C^{ab}_{L} \bar{\xi}_{b} = (C^{b}{}_{cd})_{L} \bar{\xi}_{b} =
\bar{\nabla} _{b} \left( \bar{\epsilon}_{acd} \bar{\cal{F}}^{[ab]} \right) \,. \]
In fact a close examination of \re{GCScharden} and \re{einchar}, and the way the Killing 
vector $\bar{\Xi}$ has been defined above, reveals that all is indeed consistent with \re{Yxi}.

From another perspective, our proposed current \re{jlin} would have been satisfactory if we
could write
\[ (C^{b}{}_{cd})_{L} \bar{f}^{cd} = \left( \bar{\epsilon}_{acd} \bar{f}^{cd} \right) C^{ab}_{L} 
=: C^{ab}_{L} \bar{\zeta}_{a} \,, \]
where the would-be background Killing vector \( \bar{\zeta}^{a} \), satisfying 
\( \bar{\nabla} _{(a} \bar{\zeta}_{b)} = 0 \), would let us write 
\[ (C^{b}{}_{cd})_{L} \bar{f}^{cd} = \bar{\nabla} _{a} \big( \bar{\cal{F}}^{[ab]}(\bar{\zeta}) \big) \]
on a par with \re{Yxi}. However, it is well known that 
\( \bar{\zeta}_{a} \equiv \bar{\epsilon}_{acd} \bar{f}^{cd} \) is a closed conformal
Killing vector satisfying
\[ \bar{\nabla} _{(a} \bar{\zeta}_{b)} 
= \frac{1}{3} \big( \bar{\nabla} _{c} \bar{\zeta}^{c} \big) \bar{g}_{ab} \,. \]
Hence $j^{a}$ does not lead to asymptotic AD charges.

As a final note, we would like to point out to an interesting observation we made as we were
trying out various ideas presented here using the renowned BTZ metric \cite{Banados:1992wn}
considered as a solution to TMG:
\ber 
ds^2 = \Big( M - \frac{r^2}{\ell^2} \Big) \, dt^2 
- J \, dt \, d\theta + r^2 \, d\theta^2
+ \frac{dr^2}{- M + \frac{r^2}{\ell^2} + \frac{J^2}{4 r^2}} \,.
\eer{BTZ}
The background to work with is locally AdS and obtained by setting \( M = J = 0 \) in \re{BTZ}:
\ber
d\bar{s}^2 = - \frac{r^2}{\ell^2}  \, dt^2 + r^2 \, d\theta^2 + \frac{\ell^2}{r^2} \, dr^2 \,.
\eer{bBTZ} 
The details of how to determine the energy and angular momentum of \re{BTZ} in TMG 
using the maximally symmetric background \re{bBTZ} can be found in \cite{Olmez:2005by}.
Here we just want to point out to the remarkable result that the linearized York tensor \( C^{ab}_{L} \)
that goes into the current \re{Yxi} identically vanishes in this setting, but is nevertheless able
to produce sensible \( \bar{\cal{F}}^{[ab]}(\bar{\xi}) \) as explained in \cite{Olmez:2005by}.

\section{Potentials $\&$ new mathematical relations}
\label{math}

In this section we find the potentials for the Cotton currents $j^a$ and $J^a$, uncovering 
some interesting new relations for CKYTs en route. We collect these and a relation for KYTs 
in torsionful geometries, continuing the mathematical investigations in \cite{Lindstrom:2021qrk}.

A covariantly conserved antisymmetric rank $n$ tensor represents a co-closed $n$-form. 
By the Poincar\'e lemma extended to the exterior co-derivative this means that it is equal 
to the co-derivative of an $(n+1)$-form in an appropriately chosen open set. We refer to 
this $(n+1)$-form as a potential for the conserved tensor. It should thus be possible to write 
our conserved currents $j^a$ and $J^a$ as covariant divergences of such potentials.

\subsection{A potential for $j^a$}
For $j^a$ this is straightforward using the relation
\ber
f^{ab} \, \na_a G_{bc} = 0 \,,
\eer{ide1}
recently proven in \cite{Lindstrom:2021qrk}. It implies that the current reduces to the desired 
form $j^a=\na_c\ell^{ac}$:
\ber
j^a = \left( \frac{D-2}{D-1} \right)  f^{ac} \na_c R \, 
 = \na_c \left( \left( \frac{D-2}{D-1} \right) f^{ac} R \right)  \,.
\eer{Rcot}
or, equivalently,
\ber
\tcboxmath{
\ell^{ac}=\left( \frac{D-2}{D-1} \right) f^{ac} R \,. }
\eer{jpot}

\subsection{A potential for $J^a$}
To find the potential for the Cotton current $J^a$ involving a second rank CKYT, we need to 
develop more tools.

The identity  \cite{Kastor:2004jk}
\ber
 \nabla_{a} \nabla_{b} f_{c_{1} \dots c_{n}} = (-1)^{n+1} \frac{(n+1)}{2} \,
 R^{d}\,_{a[bc_{1}} \, f_{c_{2} \dots c_{n}] d} \,, \label{ddKY}
\eer{ID11}
may be generalized to CKYTs \re{cKYT} (as well as to geometries with torsion, see below). 
From \re{cKYT}, a  second rank CKYT $k$ satisfies
\ber
\na_a k_{bc} = \na_{[a} k_{bc]} + \frac 2 {D-1} g_{a[b} \bar{k}_{c]} \,.
\eer{id1}
The modified identity corresponding to \re{ID11} then reads (see e.g.  
\cite{Batista:2014fpa}):
\ber
\nabla_{a} \nabla_{b} k_{cd} = - \frac{3}{2} R^{e}{}_{a[bc} k_{d]e} 
- \frac{3}{D-1} g_{a[b} \nabla_{c} \bar{k}_{d]}
+ \frac{2}{D-1} \nabla_{a} \left( g_{b[c} \bar{k}_{d]} \right) \,. 
\eer{LS}
We note in passing that for a closed CKYT, i.e., when the first term on the RHS of \re{id1} vanishes, 
the relation \re{LS} implies no restriction on $\bar k$.

From \re{LS} we derive the following identities
\bea
&& \na_{a} \bar k^{a} =  \na_{a} \na_{b} k^{ba} = 0 \,, \qquad 
(D-2) \, \na_{(a} \bar{k}_{b)} = (D-1) \, R^{c}{}_{(a} k_{b)c} \,, \label{LS1} \\
&& (D-4) \, \na_{[a} \bar{k}_{b]} = (D-1) \left( R^{c}{}_{[a} k_{b]c} - \BOX \, k_{ab} 
+ \half k^{cd} R_{cdab} \right) \,. \label{LS2}
\eea
Note that, when the spacetime $g_{ab}$ is an Einstein space, i.e. \( R_{ab} = K \, g_{ab} \) for 
a constant $K$, \re{LS1} shows that $\bar{k}^{a}$ is a Killing vector.
The relations \re{LS1} and \re{LS2} can be used for writing
\bea
(D-4) \, \na_{a} \bar{k}_{b} & = & (D-1) \left( \frac{D-3}{D-2} \, R^{c}{}_{a} k_{bc} 
- \frac{1}{D-2} \, R^{c}{}_{b} k_{ac} - \BOX \, k_{ab} + \half k^{cd} R_{cdab} \right) \,, \label{LS3} \\
(D-4) \, \na_{c} \na_{a} k^{c}{}_{b} & = & \frac{2D-5}{D-2} \, R^{c}{}_{a} k_{bc} 
- \frac{D-1}{D-2} \, R^{c}{}_{b} k_{ac} + \frac{3}{2} k^{cd} R_{cdab} - (D-1) \, \BOX \, k_{ab} \,.
\label{LS4} 
\eea
Furthermore, when $D = 4$, the left hand sides of \re{LS2}, \re{LS3} and \re{LS4} vanish 
and yield nontrivial constraints on $k_{ab}$. Additionally, for a flat 4-dimensional spacetime, 
$k$ has to be harmonic.

We can also perform  a calculation analogous to that of Sec.~\!\ 3.2 in \cite{Lindstrom:2021qrk}
for a rank-2 CKYT. Using \re{id1} and \re{LS} in \( [\na_a, \na_b] \na_c k_{de} \), we first find
\ber
[\na_a, \na_b] \na_{[c} k_{de]} = \frac{3}{2} \Big( \left( \na^{i} R_{ba[cd} \right) k_{e]i} 
- R^{i}{}_{b[cd} \na_{|a|} k_{e]i} + R^{i}{}_{a[cd} \na_{|b|} k_{e]i} \Big)
+ \frac{6}{D-1} g_{[c [a} \na_{b]} \na_{d} \bar{k}_{e]}.
\eer{ara}
Contracting the index pairs $(a,e)$ and $(b,c)$ in \re{ara}, we arrive at
\ber
k^{bc} \na_{c} G_{ab} + \frac{1}{D-1} \left( R \, \bar{k}_{a} + (D-4) R_{ac} \, \bar{k}^{c} 
+ 2 (D-2) \na^{c} \na_{[c} \bar{k}_{a]} \right) = 0 \,,
\eer{new}
where $G_{ab}$ is the Einstein tensor. This correctly reduces to \re{ide1}, its analog for 
a rank-2 KYT when $\bar{k}$ pieces are set to zero. Moreover, with \re{new}, the current $J^a$ 
\re{CotCur} can be expressed in an alternative form as
\ber
J^a = \frac{1}{D-1}  \left( (D-2) \, k^{ac} \na_c R
- 2 R \, \bar{k}^{a} - 2 (D-4) R^{ac} \, \bar{k}_{c} 
- 4 (D-2) \na_{c} \na^{[c} \bar{k}^{a]} \right) \,,
\eer{RJcot}
in analogy to \re{Rcot}. Using \re{RJcot}, the covariant conservation of $J^a$ can 
be verified easily thanks to the observations 
\[ 2 \na_{a} \na_{c} \na^{[c} \bar{k}^{a]} = 2 \na_{[a} \na_{c]} \na^{c} \bar{k}^{a} 
= [\na_{a}, \na_{c}] \na^{c} \bar{k}^{a} = 0 \,, \]
where the last equality follows from the definition of the Riemann tensor, and
\[ R^{ac} \na_{a} \bar{k}_{c} = R^{ac} \na_{(a} \bar{k}_{c)} = 0 \,, \]
where the last equality follows using \re{LS1}.

Note that we may alternatively write $J^a$ as
\bea
J^a & = & \na_{c} \left( \left( \frac{D-2}{D-1} \right) \left( R k^{ac} - 4 \na^{[c} \bar{k}^{a]} \right) \right)
- 2 \left( \frac{D-4}{D-1} \right) G^{ac} \bar{k}_{c} \,, \nn \\
& = & \na_{c} \left( \left( \frac{D-2}{D-1} \right) R k^{ac} + 2 G^{ab} k_{b}{}^{c} \right)
+ 2 G^{ac} \bar{k}_{c} + \left( \frac{D-2}{D-1} \right) R \bar{k}^{a} \,. \nn \\
& = & \na_{c} \left( \left( \frac{D-2}{D-1} \right) R k^{ac} + 2 G^{b[a} k_{b}{}^{c]} \right)
+ 2 \left( \frac{D-2}{D-1} \right) G^{ac} \bar{k}_{c} \,. \label{altJa}
\eea
Here the first line follows from \re{RJcot}, whereas the second one from using
\re{Cotdef} and \re{CotCur}. The current can be written as a total divergence 
as \( J^a = \na_{c} \ell^{ac} \) by equating the first and the third lines, eliminating 
the $G^{ac} \bar{k}_{c}$ bits, where
\ber
\tcboxmath{\ell^{ac} = \frac{2(D-4)}{D-3} G^{b[a} k_{b}{}^{c]} 
+ \frac{2(D-2)^2}{(D-1)(D-3)} \na^{[a} \bar{k}^{c]}
+ \left( \frac{D-2}{D-1} \right) R k^{ac}  \,.}
\eer{Japot}
Note that $k_{ab}$ becomes a KYT when $\bar k^a=0$. Agreement between \re{Japot} 
and \re{jpot} then follows by observing that 
\ber
\na_{c} \left( G^{b[a} f^{c]}{}_{b} \right) = 0
\eer{}
for a KYT $f_{ab}$ \cite{Lindstrom:2021qrk}. This also shows the consistency of
the following useful relation unveiled in the derivation of \re{Japot} 
\ber
G^{ac} \bar{k}_{c} = \left(  \frac{D-1}{D-3} \right) \na_{c} \left( 
G^{b[c} k_{b}{}^{a]} + \left( \frac{D-2}{D-1} \right) \na^{[a} \bar{k}^{c]} \right) \,.
\eer{}

\subsection{Potentials and charges for the Kerr-Newman metric}

Armed with the potential \re{Japot} we now return to the question of charges. We illustrate 
how the potential simplifies in $4D$ with vanishing curvature scalar $R=0$.  

An example of this is the Kerr-Newman metric, obtained by setting the acceleration 
$\alpha=0$ and the NUT parameter $\ell=0$ in the metric \re{newPD}, which again 
makes $\Omega=1$ and $P=1$ \cite{Podolsky:2021zwr}:
\ber
ds^{2} = - \frac{Q}{\rho^{2}} \left( dt - a \sin^{2}\theta \, d\varphi \right)^{2} 
+ \frac{\rho^{2}}{Q} dr^{2} + \rho^{2} d\theta^{2} + \frac{\sin^{2}{\theta}}{\rho^2}
\left( a dt - (r^2+a^2) \, d\varphi \right)^2 \,,
\eer{KNmet}
where we now have
\begin{eqnarray*}
Q(r) & = & \left(r-r_{+}\right) \left(r-r_{-}\right) \,, \quad \mbox{with} \quad 
r_{\pm} \equiv m \pm \sqrt{ m^{2} - a^{2}  - e^{2} - g^{2}} \,, \\
\rho^{2} & = & r^{2}+ a^{2} \cos^{2} \theta \,.
\end{eqnarray*}
We again find a KYT by setting $\alpha=0$ and $\ell=0$ in $\tilde{k}$ \re{newk} which
again leads to \( {\cal Q}_{\hat{k}} = 0 \) through $j=0$, which is also easy to see from
the potential \re{Rcot}. The charge expression ${\cal Q}_{\hat{h}}$ is easier to calculate 
using the potential $\ell^{ac}$ \re{Japot}, since now the first and the third terms trivially vanish,
for the remaining CKYT
\[ \tilde{h} = r \, dr \wedge \left[ dt - a \sin^{2}{\theta} d\varphi \right] + 
a \cos{\theta} \sin{\theta} d\theta \wedge 
\left[a \, dt - \left( r^{2} + a^{2} \right) d \varphi \right] \,. \]
We find \\
\ber{} {\cal Q}_{\tilde{h}} = 16 \pi \left(e^2+g^2\right) \int_{r_{+}}^{\infty} dr \int_{0}^{\pi} d\theta 
\frac{\sin{\theta} \left(-a^2 \cos^2{\theta} +2 a^2+e^2+g^2-2 m r+r^2\right)}{\left(a^2 \cos^2{\theta}
+r^2\right) \left(a^2 \cos^2{\theta}+e^2+g^2-2 m r+r^2\right)}
 \,, 
 \eer{KNQ}
which is certainly convergent and finite. The $\theta$ integral can be taken exactly:
\[ {\cal Q}_{\tilde{h}} = \frac{32 \pi}{a} \left(e^2+g^2\right) \int_{r_{+}}^{\infty} dr
\frac{\left( \frac{\left(2 a^2+e^2+g^2+2 r (r-m)\right)\tan^{-1}\left(\frac{a}{r}\right)}{r}-\frac{2 
\left(a^2+e^2+g^2-2 m r+r^2\right) \tan ^{-1}\left(\frac{a}{\sqrt{e^2+g^2-2 m r+r^2}}\right)}
{\sqrt{e^2+g^2+r (r-2 m)}}\right)}{\left(e^2+g^2-2 m r\right)}. \]
Unfortunately though we have not been able to evaluate the $r$ integral\rd{\footnote{Note that
in the extremal case when \( r_{+} = r_{-} = m \) and \( m^2=a^2+e^2+g^2 \), the argument of 
the $\tan^{-1}$ in the second piece at the lower limit $r=m$ becomes imaginary.}}. 
Alternatively, we can evaluate the  $r$ integral first, but that leads to an equally challenging 
expression for the $\theta$ integral.

\subsection{KYT and torsion}
\label{Torsion}
In Sec.\ref{super} the geometry is torsionful. This prompts the question of generalization of some 
of our formulae to such geometries. CKYTs with torsion have been discussed, e.g., in 
\cite{Batista:2015vxa},\cite{Papadopoulos:2011cz} and \cite{Houri:2012eq}.
Here we shall not be able to treat the general case, but limit ourselves to the special case of 
completely skew torsion. The torsionful covariant derivative is then
\ber
\hat \na=\na^{(0)}+T \,,
\eer{}
where $\na_a^{(0)}$ is the Levi-Civita connection and $T$ represents the torsion two-form
\ber
T^a=\half T_{~bc}^ae^b\wedge e^c~.
\eer{}
We consider a two-form $f$  which satisfies
\ber
\hat \na_{(a}f_{b)c}=\na^{(0)}_{(a} f_{b)c}+T_{c(a}^{~~~d} f_{b)d}=0 \,.
\eer{TKYT}
From this it follows that
\ber
\hat \na_{a}f_{bc}=\hat \na_{[a}f_{bc]}~,~~~\hat \na_{a}f^a_{~c}=0~,
\eer{TKYT2}
and we may take \re{TKYT} or \re{TKYT2} as the definition of a KY two-form in the presence 
of torsion. For a general $n$ CKYT $k$ we equivalently define
\ber
\hat\na_a k_{b_1 \dots b_{n}} = \hat\na_{[a} k_{b_1 \dots b_{n}]} 
+ \frac{n}{D-n+1} \, g_{a[b_1} \, \bar{k}_{b_2 \dots b_{n}]} \,,
\eer{CKYT}
where
\ber
\bar k_{b_1 \dots b_{n-1}} := \hat\na_c k^c_{~b_1 \dots b_{n-1}}~.
\eer{}
For a KY two-form $f$, we then find, using the torsionful Bianchi identity
\ber
\hat R_{[dbc]}{}^{e}=\na_{[d} T_{bc]}^{~~e}+T_{[db}^{~~f}T_{c]f}^{~~e} \,,
\eer{TBian}
that the  identity corresponding to \re{ID11} becomes
\ber
 \hat\na_a \hat \na_{b} f_{cd} =  {\textstyle \frac{3}{2}} \,
 \hat R\,_{[bc~|a|}^{~~~e} \, f_{d] e}+{\textstyle \frac{3}{2}}\left( \hat \nabla_{[d} T_{bc]}^{~~e}+T_{[db}^{~~f}T_{c]f}^{~~e}\right)f_{ea}-4\hat R_{[abc}{}^{e}~\!f_{d]e}~.
\eer{ID2}
The explicit torsion expression for the last term follows from \re{TBian}.

\section{Conclusions}
\label{conc}
In this paper we have introduced conserved currents based on the Cotton tensor and second 
rank (C)KYTs and derived their potentials. As an example we have studied the 
Pleba\'nski-Demia\'nski metric and found its two CKYTs in the new coordinates 
introduced in \cite{Podolsky:2021zwr}. The corresponding Cotton charges are proportional 
to the sum of the squares of the electric and the magnetic charges in general. In the particular 
case of the charged Taub-NUT spacetime, which corresponds to setting 
the rotation parameter $a$ to zero, the charge for the non-vanishing CKYT is simply 
$32\pi^2(e^2+g^2)/\ell$ with the NUT parameter $\ell$. Obviously, the conserved charge 
we find is a combination of fundamental charges already present in the metric. The existence 
of KYTs and conserved currents formed using them does {\em not} automatically lead to new 
charges. This is evident in the case of black holes where all the conserved charges are known 
by other means. However for other possibly exotic geometries, new and hitherto unearthed 
(hidden) charges could well arise. Therefore, there are other metrics waiting to be 
analyzed similarly, in particular in higher dimensions. 

For the $3D$ case  we have discussed possible asymptotic charges, relations to results in 
TMG and constructed a generalization to conformal supergravity. This latter result opens 
up the possibility to construct similar currents for supergravities in higher dimensions. 

The derivation of the potentials for the Cotton currents involve generalizations of 
identities for KYTs and CKYTs, and shows that there are still interesting 
geometric relations for these objects that deserve attention.\\

\noindent{\bf Note added in proof:} While this paper was a preprint, we were asked 
by D. Kubiznak what would happen if one considered the $\ell = 0$ case, i.e., the
Reissner-Nordstr\"om case in the example leading to \re{theQ}. The whole
discussion follows as before, but the range of the integral in  \re{theQ} is 
from $r_+$ to $\infty$ now, and one has a well-defined finite answer.\\

\noindent{\bf Acknowledgments}\\
We thank Sergei Kuzenko for commenting on the supergravity part of the paper,
and D. Kubiznak for comments. The research of U.L. is supported in part by the 
2236 Co-Funded Scheme2 (CoCirculation2) of T\"UB{\.I}TAK 
(Project No:120C067)\footnote{\tiny However the entire responsibility for the publication is ours. 
The financial support received from T\"UB{\.I}TAK does not mean that the content of the 
publication is approved in a scientific sense by T\"UB{\.I}TAK.}.\\

\centerline{\Large \bf Appendix}
\appendix
\section{Spinor algebra}
\label{Spinor}
The $3D$ spinor algebra we need is
\ber\nn
&&\{\ga_a,\ga_b\}=2\eta_{ab} \,, \\[1mm]\nn
&&\epsilon^{\al\be}\epsilon_{\be\ga}=\de^\al_\ga~,~~~\psi^\al=\epsilon^{\al\be}
\psi_\be~,~~~\psi_\al=\epsilon_{\al\be}\psi^\be  \,, \\[1mm]\nn
&&\psi_\al\phi_\be-\psi_\be\phi_\al=\epsilon_{\al\be}\psi^\ga\phi_\ga  \,, \\[1mm]\nn
&&\psi^\al\phi^\be-\psi^\be\phi^\al=-\epsilon^{\al\be}\psi^\ga\phi_\ga  \,, \\[1mm]\nn
&&\ga_a\ga_b=\eta_{ab}+\epsilon_{abc}\ga^c \,, \\[1mm]\nn
&&(\ga^a)_{\al\be}(\ga_a)_{\ga\de}=2\epsilon_{\al(\ga}\epsilon_{\de)\be} \,, \\[1mm]\nn
&&V_{\al\be}=(\ga^a)_{\al\be}V_a=V_{\be\al} \,, \\[1mm]
&&V_a=-\half (\ga_a)^{\al\be}V_{\al\be}  \,.
\eer{}
Note that a vector index is equivalent to a symmetric pair of spinor indices through the last relation.
\ber
\eta_{ab}=\eta_{(\al\be)(\ga\de)}=-\epsilon_{\al(\ga}\epsilon_{\de)\be}  \,.
\eer{}

\end{document}